# Volatile $MoS_2$ Memristors with Lateral Silver Ion Migration for Artificial Neuron Applications

Sofia Cruces[1], Mohit D. Ganeriwala[2], Jimin Lee[1], Lukas Völkel[1], Dennis Braun[1], Annika Grundmann[3], Ke Ran[4,5,8], Enrique G. Marín[2], Holger Kalisch[3], Michael Heuken[3,6], Andrei Vescan[3], Joachim Mayer[4,5], Andrés Godoy[2], Alwin Daus[1,7,*] and Max C. Lemme[1,8*].

[1] Chair of Electronic Devices, RWTH Aachen University, Otto-Blumenthal-Str. 25, 52074 Aachen, Germany

[2] Department of Electronics and Computer Science, Universidad de Granada, Avenida de la Fuente Nueva S/N 18071, Granada, Spain

[3] Compound Semiconductor Technology, RWTH Aachen University, Sommerfeldstr. 18, 52074 Aachen, Germany

[4] Central Facility for Electron Microscopy, RWTH Aachen University, Ahornstr. 55, 52074, Aachen, Germany

[5] Ernst Ruska-Centre for Microscopy and Spectroscopy with Electrons (ER-C 2), Forschungszentrum Jülich GmbH, Wilhelm-Johnen-Str., 52425 Jülich, Germany

[6] AIXTRON SE, Dornkaulstr. 2, 52134 Herzogenrath, Germany

[7] Sensors Laboratory, Department of Microsystems Engineering, Georges-Köhler-Allee 103, 79110 Freiburg, Germany

[8] AMO GmbH, Advanced Microelectronic Center Aachen, Otto-Blumenthal-Str. 25, 52074 Aachen, Germany

## Abstract

Layered two-dimensional (2D) semiconductors have shown enhanced ion migration capabilities along their van der Waals (vdW) gaps and on their surfaces. This effect can be employed for resistive switching (RS) in devices for emerging memories, selectors, and neuromorphic computing. To date, all lateral molybdenum disulfide ($MoS_2$)-based volatile RS devices with silver (Ag) ion migration have been demonstrated using exfoliated, single-crystal $MoS_2$ flakes requiring a forming step to enable RS. Here, we present volatile RS with multilayer $MoS_2$ grown by metal-organic chemical vapor deposition (MOCVD) with repeatable forming-free operation. The devices show highly reproducible volatile RS with low operating voltages of approximately 2 V and fast switching times down to 130 ns considering their micrometer-scale dimensions. We investigate the switching mechanism based on Ag ion surface migration through transmission electron microscopy, electronic transport modeling, and density functional theory. Finally, we develop a physics-based compact model and explore the implementation of our volatile memristors as artificial neurons in neuromorphic systems.

## 1. Introduction

Two-dimensional (2D) layered materials, including semiconducting transition metal dichalcogenides (TMDCs), have long been known to allow ion intercalation and migration within their van der Waals (vdW) gaps and on their surfaces [1–13]. More recently, these unique properties have been utilized to demonstrate applications such as emerging memory technologies, electronic switches (e.g. as selectors), and neuromorphic computing [8,10,14–21]. In certain device configurations, the number and type of defects and specific intrinsic properties of layered 2D materials (2DMs) have been shown to influence device performance and variability [13,22–26]. In the case of resistive switching (RS), the presence of defects such as cracks, lattice deformations, and dopants can positively impact RS, e.g. lower the switching voltages [27,28]. This is in contrast to 2DM-based field effect transistors, which require nearly defect-free 2DMs to ensure high charge carrier mobility, low charge trap density, structural stability, and resistance to degradation over time [28,29]. Hence, understanding the growth of large-scale layered 2DMs is important for delivering application-specific materials for future integration into semiconductor fabrication lines [27-29].

Previously, RS devices based on lateral ion migration on smooth molybdenum disulfide ($MoS_2$) surfaces have been reported [7,10,12,17,33]. However, these approaches are limited to nonscalable device configurations employing mechanically exfoliated $MoS_2$ or single-crystal $MoS_2$ flakes and face scalability, uniformity, and variability issues [7,10,12,17,33]. In addition, metal-organic chemical vapor deposition (MOCVD) or chemical vapor deposition (CVD) have been used to grow large-area 2DM films [31-33]. Thus far, lateral devices with memristive behavior induced by vacancy or grain boundary migration in CVD-grown monolayer $MoS_2$ have been studied. However, the voltages applied to switch these devices to their on-state were between 3.5 V and 20 V, which is too high for low-power applications, such as neuromorphic computing [37,38]. Moreover, these lateral $MoS_2$ devices required “forming” to initiate RS, i.e., the application of large voltages and currents to induce the first switching event. These factors can induce damage or lead to high device-to-device variability [20,39]. In addition, neuromorphic computing concepts typically involve integration with advanced complementary metal oxide semiconductor (CMOS) circuits, which cannot easily provide such high voltages. Moreover, the potential of lateral 2DM-based resistive switches as artificial neurons remains largely unexplored [6,18].

Here, we study volatile 2DM-based lateral memristive devices operating through silver (Ag) ion migration with low operating voltages and high reproducibility, including their potential as artificial neurons. We demonstrate forming-free, volatile RS in micron-sized memristive devices fabricated with multilayer $MoS_2$ grown by MOCVD [36,40,41]. We explore the current–voltage (*I-V*) characteristics and switching dynamics of devices with lateral distances between electrodes (gaps) down to ~1 µm. Furthermore, we present highly reproducible volatile RS with over several hundred cycles, fast switching down to 130 ns, and low switching voltages of ~2 V, which is a notable advancement for micron-sized lateral 2DM-based memristive devices [37,38,42]. The experimental realization is complemented with a physics-based model supported by *ab-initio* simulations. The model connects the migration dynamics of Ag ions with electron transport along our palladium (Pd)/$MoS_2$/Ag structures. Finally, we use the model to show the potential operation of our $MoS_2$ volatile memristors as artificial neurons in neuromorphic systems.

## 2. Results and Discussion

The devices investigated in this work consisted of multiple layers of MOCVD-grown $MoS_2$ with one Pd and one Ag electrical contact. The Ag ions can migrate on the surface of the $MoS_2$ layers and modulate the channel conductivity, driven by a lateral electrical field. Figure 1a shows a schematic of such a lateral $MoS_2$ memristive device.

$MoS_2$ was grown via MOCVD on 2" sapphire and wet transferred onto a $SiO_2$/Si substrate (see Experimental Section). The metal contacts were deposited by electron-beam (e-beam) evaporation of 50 nm Pd and 50 nm Ag. An additional 50 nm aluminum (Al) layer was evaporated on top of the Ag contacts to avoid Ag tarnishing [43] (see Supporting Information Figure S1). The $MoS_2$ channels in the gap between contacts were patterned via reactive ion etching (RIE), and afterwards, the complete removal of $MoS_2$ next to the channels, was verified via Raman mapping (Figure S2, Supporting Information). Figure 1b shows a top-view optical microscopy image of a fabricated device with a gap size of 5 µm. The experiments included devices with gaps between ~1 µm and ~6 µm. Figure 1c displays a top-view scanning electron microscopy (SEM) image of the polycrystalline structure of $MoS_2$ after device fabrication. Two different features can be distinguished: white triangles (marked in red), which are probable multilayer nucleation sites, and brighter white features, which may be vertically aligned sheets from the growth process. The images do not show photoresist residues on the $MoS_2$ surface from the fabrication process. High-resolution transmission electron microscopy (HRTEM) cross-section images of the Pd and Ag/Al top contacts on $MoS_2$ are presented in Figure 1d. The layered 2D structure of the multilayer $MoS_2$ film can be clearly recognized, with an interlayer distance of 6.2 ± 0.05 Å, which coincides with reported values in the literature [11,44]. Figure 1e presents an atomic force microscopy (AFM) image of the polycrystalline, as-grown $MoS_2$ on 2” sapphire. Figure 1f depicts the morphology of the material after transfer, and the inset height profile indicates a thickness of ~6 nm. HRTEM cross-sectional images at two different positions on the Pd electrode side revealed slight thickness variations in the $MoS_2$ of approximately 5 layers (Figure S3, Supporting Information). In addition, we performed Raman analysis of the as-grown material on sapphire and after transfer. The measurements were not taken at the exact same location, which could lead to a slight frequency difference due to thickness variation in the sample [45]. The extracted $E^1_{2g}$ and $A_{1g}$ peaks of $MoS_2$ coincide with the literature values for more than 4 layers or the bulk, which is in agreement with the AFM and HRTEM data (see Supporting Information Figure S4) [45,46].

We performed *I-V* measurements of the lateral $MoS_2$-based memristors by applying a voltage to the Ag electrode while the Pd electrode and the back gate were grounded. Figure 2a presents ten subsequent voltage sweeps in both positive and negative polarities that show similar volatile RS behavior. The *I-V* sweeps were conducted first in the positive direction from 0 V to 5 V (arrow number 1) and back to 0 V (arrow number 2), followed by the same procedure in the negative direction. This first switching cycle is marked in blue to show that the initial RS required no forming event at higher voltages. This behavior was specifically observed for devices with gap sizes ranging between ~1 µm and ~2 µm. Figure 2b displays over 416 switching cycles on a device with a gap size of approximately 1.2 µm. Initially, the device is in a high-resistance state (HRS) during the forward sweep until the transition to a low-resistance state (LRS) occurs at an “on-threshold” voltage ($V_{t,on}$) of approximately 2.1 V. For this measurement, the current compliance (CC) was set to 1 µA. The device remained in the LRS during the backward sweep until it switched back below a “hold” voltage ($V_{hold}$) of approximately 1.7 V. The original HRS was reached at an “off-threshold” voltage ($V_{t,off}$) of approximately 0.2 V. Moreover, we observed volatile RS over several cycles at a comparatively higher CC (see Figure S5, Supporting Information).

Furthermore, we conducted *I-V* sweeps on four devices with different metal combinations with and without $MoS_2$ to support our hypothesis that the RS originates from Ag ion migration on or between the $MoS_2$ layers. We observed volatile RS only for the Pd/$MoS_2$/Ag device (see Figure S6, Supporting Information).

We plotted $V_{t,on}$, $V_{t,off}$ and $V_{hold}$ of the 416 switching cycles in a histogram as an indirect measure of endurance and fitted the data with a Gaussian distribution (Figure S7, Supporting Information). The statistical distribution of the characteristic voltages shows reasonably low standard deviations ($\sigma$) of ~0.1 V and low cycle-to-cycle variabilities of $V_{t,on}$ and $V_{hold}$ of 4.76 % and 5.95 %, respectively. Figure 2c presents the normalized cumulative distribution functions (CDFs) of $V_{t,on}$, $V_{hold}$, and $V_{t,off}$ of the 416 switching cycles, confirming the low variability among them. We attribute this low variability to the polycrystalline nature, thickness variations, and specific topography of our MOCVD $MoS_2$ layers, in combination with our micron-sized devices [37,38]. Since the numbers of grain sizes and thickness variations are high in each device, the variability in current transport is reduced because the individual differences in the resistance average out [47]. The dependence of $V_{t,on}$ upon the gap size between electrodes is plotted in Figure 2d, with all subsequent *I-V* s weeps for each device displayed in Supporting Information Figure S8. The devices with gap sizes under 2.1 µm did not require a forming step, as previously mentioned. Devices with gap sizes of 3.9 µm and above, in contrast, required larger initial voltages between 10 and 50 V, depending on the gap size, to exhibit RS behavior.

The forming-free switching of the 1 µm and 2 µm devices is strikingly different from that in prior works using single-crystal exfoliated $MoS_2$, where forming was always necessary, and switching occurred only in submicron gaps [7,12]. Additionally, as illustrated in Figure 2d, the mean $V_{t,on}$ for the device with a 1.2 µm gap size is approximately 2.1 V. This value is mostly lower than or at least equal to the reported $V_{t,on}$ values for devices with similar or shorter gap sizes [7,33,37,38,48,49]. A benchmarking table with more parameters is available in Table S1 in the Supporting Information.

We analyzed the switching dynamics to characterize the transient behavior of our devices. Thus, we studied the time-dependent current-response of the devices upon application of voltage pulses with various amplitude and timing parameters [50,51]. The switching time ($t_{on}$) was defined as the time needed to reach 90% of the ON current ($I_{on}$), and the recovery time ($t_{off}$) was defined as the time required to reach 10% of the difference between $I_{on}$ and the OFF current ($I_{off}$) when switching back to the HRS. More information about how the switching parameters were extracted can be found in Figure S9 (see Supporting Information). Figure 3a displays a typical time-dependent response upon the application of a 5 V pulse with a duration of 5 µs to a device with a 1.2 µm gap size. We obtained characteristic response times, such as $t_{on}$ and $t_{off}$, of <400 ns and <150 ns, respectively. $t_{on}$ and $t_{off}$ are crucial parameters for memristive device operation. Both switching parameters can be influenced by external factors such as the voltage pulse amplitude and the width of the pulsed waveform. For example, with the application of a voltage pulse of 2 V and length of 20 ms, the extracted values for $t_{on}$ and $t_{off}$ drastically change to 1.2 ms and 1.45 µs, respectively (Figure 3b). Furthermore, we applied ten consecutive pulses with varying pulse widths and increasing voltage amplitudes to extract the ON-state resistance ($R_{on}$), $t_{on}$ and $t_{off}$ for each switching cycle. Table S2 summarizes the pulse programming parameters and is available in the Supporting Information. Figure 3c shows how $R_{on}$ decreases with increasing voltage pulse amplitude. The increasing voltage leads to greater Ag ion diffusion, which decreases the average distance between the ions and increases the current (see model in Supporting Information Section S13). We analyzed the switching dynamics of $t_{on}$ and $t_{off}$ in more detail (Figure 3d) and found that $t_{on}$ decreases exponentially with increasing voltage pulse amplitude. This behavior is consistent with other ion diffusion-based memristive devices [52–54]. Moreover, for the 4 V and 5 V pulse amplitudes, $t_{on}$ displays the lowest variability and the fastest switching times. It has been shown in the literature that nanoscale 2DM-based memristive devices usually have switching times in the nanosecond range (30 ns to 200 ns) [14,19,24]. Here, we observe such fast switching in lateral $MoS_2$-based memristive devices with micron-sized gaps. We attribute this behavior to the presence of grain boundaries in our polycrystalline $MoS_2$, which may facilitate Ag ion migration in the lateral structures [37,38,55]. After several *I–V* sweeps, Ag ions are distributed across the $MoS_2$ channel, thereby reducing the effective distance between electrodes and resulting in faster switching times [12,27,37,55]. The mean $t_{on}$ of our micron-sized gap devices for 4 V and 5 V ranges from 130 ns up to 200 ns. Such fast-switching performance would be beneficial for selector and neuromorphic computing applications [52,53,56,57]. In the case of 2 V and 3 V pulses, the device shows higher cycle-to-cycle variability and longer switching times ranging from 0.2 ms up to 4.5 ms. In addition, the devices required longer pulses in the millisecond range to observe the RS. Nevertheless, slower switching times at low voltages could be advantageous for some applications. For example, recent works have shown that threshold memristors with low voltages and long switching times are a natural choice for the realization of artificial neurons [6,10,58].

We analyzed the current conduction mechanisms in the HRS and LRS using the median of the *I*-*V* cycles shown in Figure 2a and found different transport mechanisms in the HRS and LRS. In the HRS (i.e., for forward sweep voltages ranging up to ~$V_{t,on}$), the device shows a linear relation between ln(*I*/*V*) and sqrt(*V*) (Figure 4a). Such a dependency is indicative of hopping transport through localized states (such as Poole–Frenkel (PF) hopping), where the mobility is increased due to the applied electric field [59–61]. In the LRS, in contrast, the *I*-*V* characteristics best match the space-charge-limited conduction (SCLC) mechanism (Figure 4b). In particular, the ln(*I*) vs ln(*V*) plot (for the backward sweep starting at the CC down to voltages between $V_{hold}$ and $V_{t,off}$) shows a slope of ~2.7, indicating that the current in the LRS follows the Mott–Gurney law of SCLC with PF-enhanced mobility ($I \propto V^{2+m}, m \geq 0$) [62]. The current conduction mechanism, therefore, changes from PF hopping to SCLC during the voltage sweep of memristive devices. This switching can be attributed to the movement of Ag ions along the $MoS_2$ surface, which have been shown to be highly active and responsible for RS in previous investigations [7,12,52] (see insets in Figure 4a and 4b). The dependence of $t_{on}$ on the applied pulse amplitude (Figure 3d) also points toward the involvement of ionic movement in the RS.

We carried out first-principles simulations using density functional theory (DFT) to analyze the electronic properties of $MoS_2$ in the presence of Ag atoms (see the Experimental Section for more details). In particular, Figure 4c shows the calculated band structure of monolayer $MoS_2$ with a low concentration of Ag atoms adsorbed on the surface on a highly symmetric path along the Brillouin zone (marked in blue), together with the band structure of a pristine $MoS_2$ layer (marked in red) for comparison. The adsorption of the Ag atom on the $MoS_2$ surface gives rise to a state in the bandgap close to the conduction band. However, the near dispersion-less nature of this state indicates that it is spatially localized. Such a state could facilitate the hopping of a carrier through the otherwise semiconducting $MoS_2$, as depicted schematically in the inset of Figure 4a. Increasing the surface concentration of Ag atoms reduces their average distance, resembling the formation of a conductive filament and creating a noticeable change in the Ag energy states (Figure 4d). The localized states are thus transformed into a delocalized band, which supports band-like transport. The intermediate stages of this transformation correspond to the gradual increase in the Ag concentration and the associated transformation of the band gap state from a localized to a delocalized band is detailed in Supporting Information Figure S10.

On the basis of these results, the RS mechanism can be understood as follows: initially, the memristor is in the HRS. The application of a voltage bias then results in the drift of active Ag ions across the $MoS_2$ surface, causing carrier transport by PF hopping through the localized, Ag ion-induced states. The polycrystalline $MoS_2$, which was used in this work, may also exhibit similar PF hopping conduction, even when the Ag concentration is negligible [63]. Increasing the voltage bias increases the Ag ion concentration, which modifies the band structure of the material and enables band-like transport, leading to higher current levels in the LRS. Note, that the Ag ions may not form a continuous conductive filament. If the concentration of the Ag ions is high enough, the average distance between them reduces, giving rise to a de-localized band, thereby changing the transport mechanism. The volatile nature of the RS and the non-ohmic *I-V* relationship in the LRS in our devices can be explained by the formation of a discontinuous Ag filament, which self-ruptures when the voltage bias is reduced below $V_{hold}$ as the Ag ions diffuse away, resetting the device back to the HRS [52].

Our volatile lateral $MoS_2$ memristors, with their dynamic thresholding, can be harnessed to implement artificial neurons in neuromorphic computing systems [6,64]. To demonstrate this possibility, we developed a physics-based model capturing the observed Ag dynamics that give rise to the RS, as well as the electronic current conduction mechanisms in both resistive states. Details of the model and its equations are summarized in section S13 in the Supporting Information. Figure 5 shows the *I*-*V* model, which phenomenologically captures the experimental dynamics of the volatile memristors to a high degree. This enables us to emulate the experimental behavior of the memristor in circuit-level simulations. We implemented the model in Verilog-A and investigated the leaky integrate-and-fire (LIF) model of the neuron shown in Figure 5b. Here, the capacitor (*C*), along with the parallel memristor, operates as an LIF neuron's soma, whereas the series resistance (*R*) acts as the synaptic weight. The SPICE simulation results (Figure 5c) show the input pulse voltage ($V_{in}$) emulating the spike signals arising from the presynaptic neurons. Since the parallel memristor is initially in its HRS, the capacitor integrates the input signal into its membrane potential ($V_c$). When the potential reaches the threshold voltage of the neuron (~$V_{t,on}$), the memristor switches to the LRS, resulting in a sudden current spike (Figure 5d) equivalent to the firing of a neuron. The LRS of the memristor enables the discharging of the capacitor, mimicking leaky neuron behavior, and $V_c$ reaches its original potential after some period of inactivity of the presynaptic neuron, as shown in Figure 5d. After firing, the cell must wait for the so-called refractory period before it can begin a new integration and firing process [65]. Our circuit-level simulations demonstrate the suitability of the fabricated $MoS_2$ lateral memristors as artificial neurons, confirming their potential for application in future neuromorphic systems.

## 3. Conclusion

We demonstrated forming-free, volatile RS in memristive devices with micrometer-sized gaps fabricated from multilayer polycrystalline $MoS_2$. Our devices exhibit highly reproducible volatile RS with over several hundred cycles, switching voltages of ~2 V, and fast switching within a few hundred nanoseconds for 5 V pulses. We analyzed the current conduction mechanisms for both the HRS and LRS and concluded that Poole-Frenkel hopping and space-charge limited conduction are dominant, respectively. In addition, we captured the switching dynamics of Ag ions and explained the electron transport along our experimental lateral Pd/$MoS_2$/Ag-Al memristors with a physics-based model. Finally, the model was employed to demonstrate the potential of our volatile memristors for artificial neurons in neuromorphic systems.

## 4. Experimental Section

**Metal–organic chemical vapor deposition (MOCVD) of $MoS_2$:** Highly uniform $MoS_2$ was epitaxially grown in a commercial AIXTRON planetary reactor in a 10×2" configuration on sapphire (0001) substrates. First, the substrate was prebaked at 1050 °C in a pure $H_2$ atmosphere to promote lateral growth by preventing the formation of a parasitic carbonaceous film[36]. The growth process was carried out at a substrate temperature of 845 °C, with nitrogen as the carrier gas and a pressure of 30 hPa. A high sulfur to molybdenum ratio of 200000 was chosen to achieve homogeneous $MoS_2$ films on a wafer scale. To achieve this, the precursor flow rate was set to 0.1 nmol/min for molybdenum hexacarbonyl (MCO) and 20 µmol/min for di-tert-butyl sulfide (DTBS) [40].

**Device Fabrication:** $MoS_2$ grown on 2" sapphire by MOCVD was transferred onto 2 x 2 $cm^2$ Si chips covered with 275 nm thermal $SiO_2$. Poly(methyl methacrylate) (PMMA) was spin-coated on top of the $MoS_2$ before being released from the sapphire substrate in a potassium hydroxide (KOH) solution [40]. The electrodes were defined with the AZ5214E JP photoresist from Merck Performance Materials GmbH and optical contact lithography with an EVG 420 Mask Aligner. The asymmetric Pd (50 nm) and Ag (50 nm)/Al (50 nm) electrodes were deposited via electron-beam evaporation in a Pfeiffer tool and subsequently lifted in acetone at room temperature. Finally, the channels were patterned via $CF_4/O_2$ reactive ion etching (RIE) in an Oxford Instruments Plasma Lab System 100 tool.

**Material and Device Characterization:** Optical microscope images were recorded with a Leica INM100 microscope and a Keyence Laser Scanning microscope. Raman measurements were performed with a WiTec alpha300R Raman spectrometer in mapping mode with an excitation laser wavelength of 532 nm and 1 mW laser power. Scanning electron microscopy (SEM) images were taken with a Zeiss Supra 60VP SEM at an operation voltage of 4 kV. Transmission electron microscopy (TEM) analysis was conducted with a JEOL JEM F200 instrument at 200 kV. The lamella was prepared via a focused ion beam (FIB) with an FEI Strata400 system with a gallium (Ga) ion beam. Atomic force microscopy (AFM) measurements were conducted with a Dimension Icon AFM from Bruker Instruments in tapping mode.

**Electrical Measurements:** Electrical measurements were performed in a LakeShore probe station connected to a semiconductor parameter analyzer (SPA) "4200A-SCS" with two source measure unit (SMU) cards "Keithley 4200-SMU", each connected to a preamplifier "Keithley 4200-PA" from Tektronix. A voltage was applied to the Ag/Al electrode, and the Pd electrode was grounded. *Current–voltage (I–V)* measurements were conducted by sweeping the voltage from 0 V to a positive maximum voltage $V_{max}$ and back to 0 V. Cycling measurements were performed with a 0 s delay between each sweep cycle in normal mode. The current is limited by an external current limiter within the semiconductor parameter analyzer. Pulse experiments were performed by supplying a voltage to the Ag/Al electrode (channel 1) and measuring the output current over time in channel 2 (Pd electrode).

**Theoretical Calculations:** Density functional theory (DFT) calculations were carried out via the generalized gradient approximation (GGA) as implemented in QuantumATK with a linear combination of atomic orbitals (LCAO) basis set using the Perdew-Berke-Erzenhof exchange-correlation functional [66,67]. An SG15 pseudopotential with a medium-size basis set was selected for the simulations [67]. A vacuum layer of 20 Å was fixed along the out-of-plane direction to avoid interactions with periodic replicas of the system. First, a 5x5 supercell with a single Ag atom (corresponding to a concentration of $4x10^{13}cm^{-2}$) was considered. To further increase the Ag concentration, the supercell size was gradually reduced from 5x5 to 2x2. The Brillouin-zone integration was performed over a Monkhorst-Pack grid of k-points with a density (Å) of 4x4x1. A sufficiently large energy cutoff of 150 Ry was considered. The van der Waals (vdW) interactions were considered in the calculations through Grimme's DFT-D2 dispersion corrections. The geometries and lattice parameters were fully relaxed until the forces acting on each atom were less than 0.01 eV/Å. The integrity of the simulations was first verified by comparing the band structure of the pristine monolayer $MoS_2$ and the obtained DFT-based band gap of ~1.7 eV with the literature [68].

**Acknowledgments**

Financial support from the German Federal Ministry of Education and Research (BMBF) within the projects NEUROTEC 2 (No. 16ME0399, 16ME0400, and 16ME0403) and NeuroSys (No. 03ZU1106AA and 03ZU1106AD) is gratefully acknowledged. M. D. Ganeriwala acknowledges funding from the European Union's Horizon 2020 research and innovation program under the Marie Sklodowska-Curie grant agreement No. 101032701. This work was also supported by the Spanish Government through research projects PID2020-116518GBI00 funded by MCIN/AEI/10.13039/501100011033 and TED2021-129769B-I00 funded by MCIN/AEI/10.13039/501100011033 and the European Union NextGenerationEU/PRT.

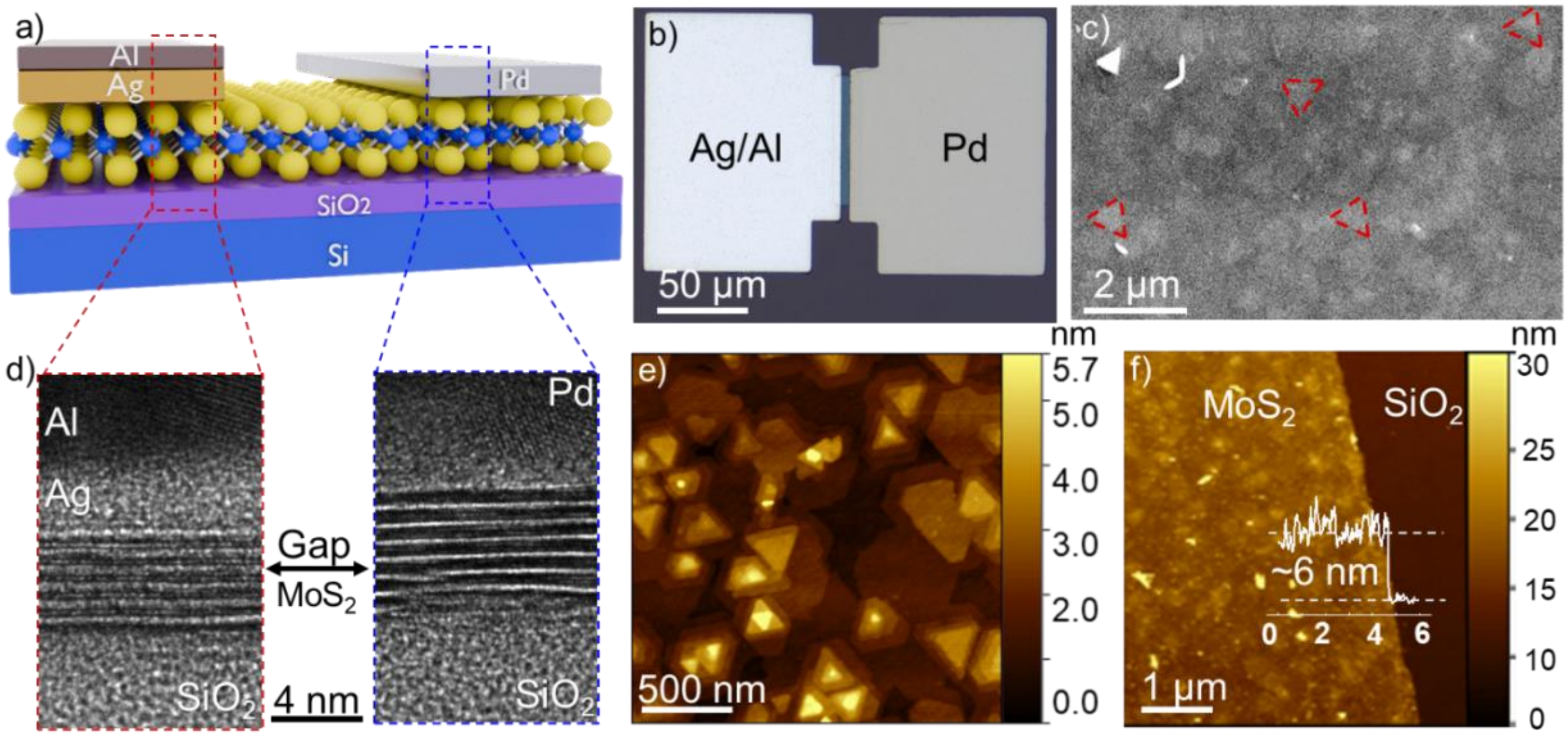

**Figure 1**. Cross-section of the device structure and material characterization data. a) Schematic and b) optical microscopy image of our Al/Ag/$MoS_2$/Pd device. c) Top-view SEM image of the crystalline structure of $MoS_2$ after device fabrication. The white triangles (marked in red) indicate possible multilayer nucleation sites. d) TEM images of both electrodes and $MoS_2$. e) AFM image of MOCVD $MoS_2$ as grown on 2” sapphire. g) AFM image of $MoS_2$ after transfer to $SiO_2$/Si. The inset height profile shows a thickness of ~6 nm.

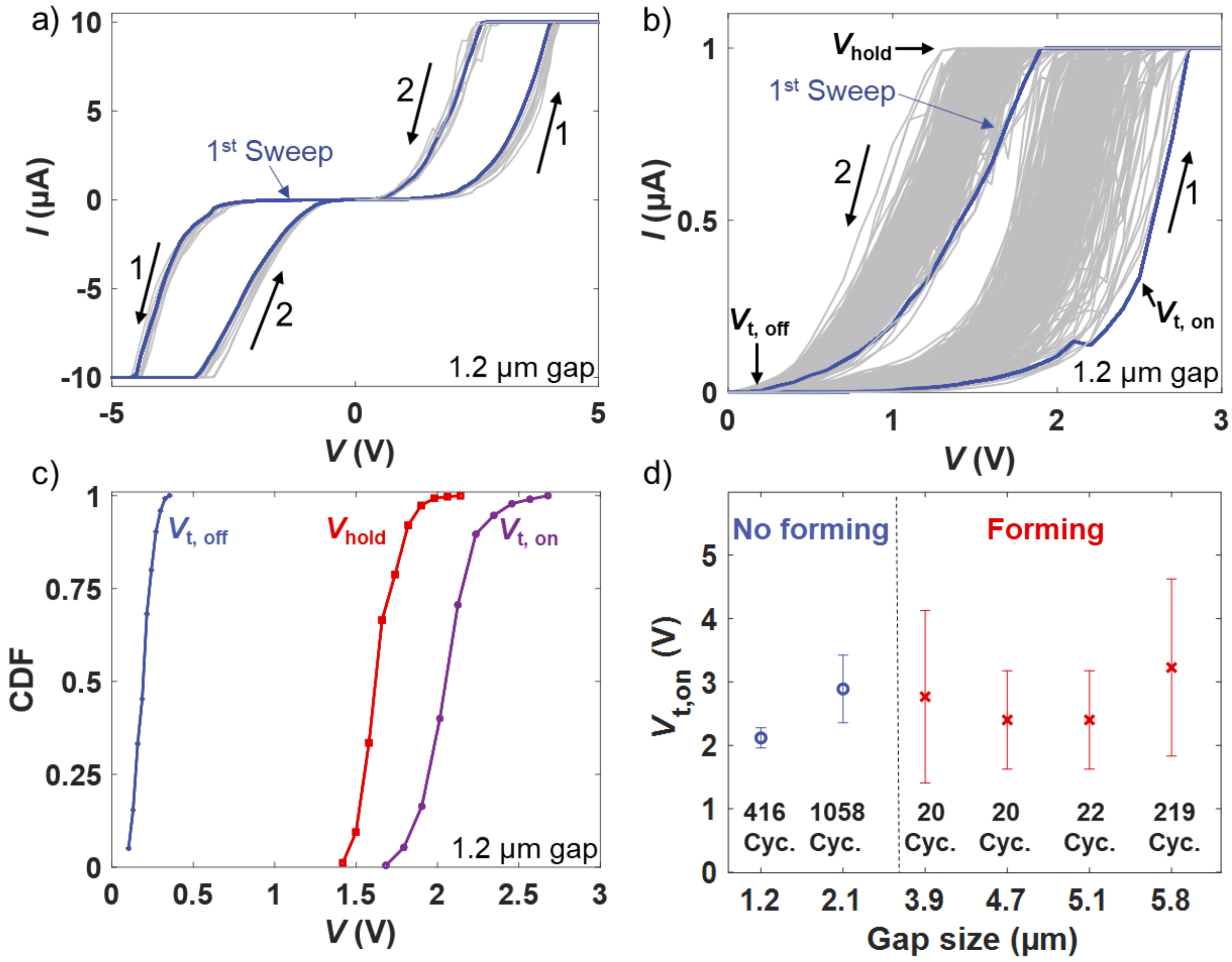

**Figure 2**. DC *I-V* characterization of Pd/$MoS_2$/Ag memristive devices. a) 10 *I–V* curves recorded in both voltage polarities. Arrows 1 and 2 show the voltage sweep direction. The first sweep is marked in blue. b) Over 400 consecutive switching cycles from a 1.2 µm gap device. The characteristic voltages are marked on the curves. c) Cumulative distribution function of $V_{t,on}$, $V_{t,off}$ and $V_{hold}$ of the data displayed in b). d) Dependence of $V_t$ on the gap size for different devices with different numbers of subsequent RS cycles.

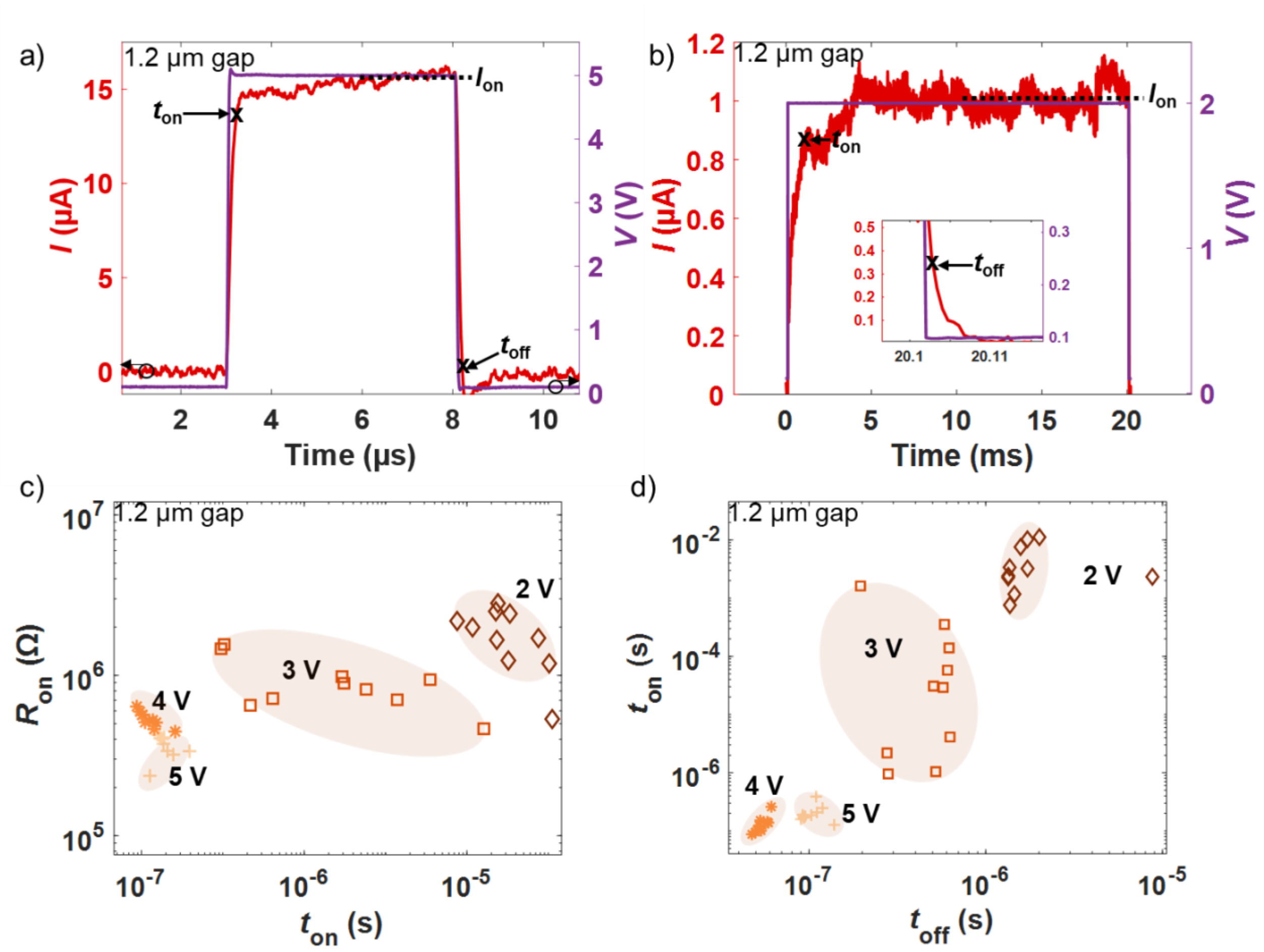

**Figure 3.** Dynamic response of threshold switching in an ~1.2 µm gap size device. a) Pulsed waveform at a 5 V pulse amplitude showing switching on the order of 400 ns. b) Pulsed waveform at a 2 V pulse amplitude showing volatile switching on the order of 1.2 ms. c) Cycle-to-cycle variability of the ON-state resistance ($R_{on}$) and switching time ($t_{on}$) for ten sequentially pulsed waveforms at different pulse amplitudes. The device has a larger $t_{on}$ with decreasing voltage pulse amplitude. d) Correlation of $t_{on}$ and recovery time ($t_{off}$) at different pulse amplitudes with varying pulse width at which RS was observed.

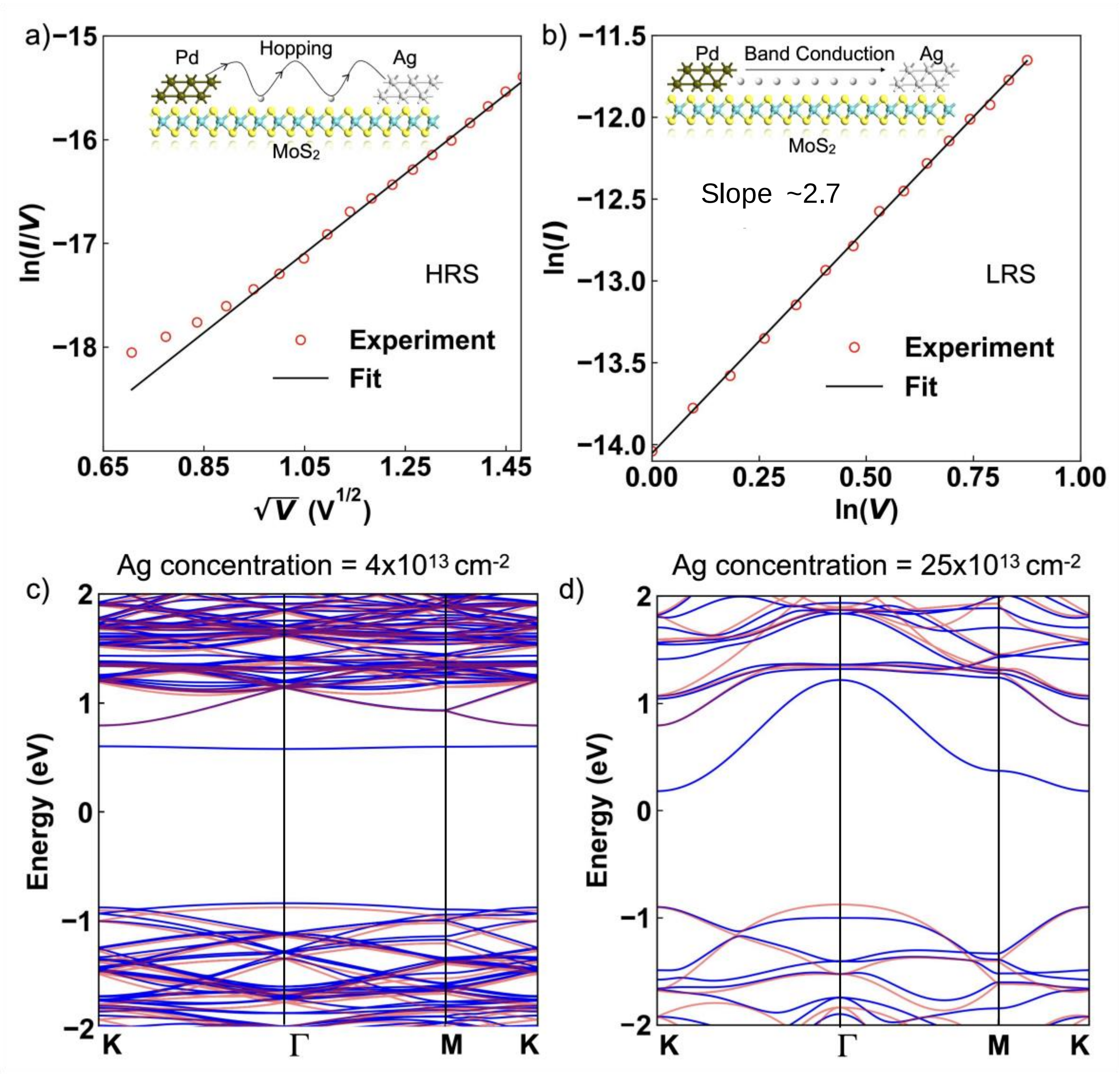

**Figure 4.** *I-V* characteristics of the volatile memristor. a) HRS shows Poole–Frenkel-type hopping transport, while b) LRS shows space-charge limited conduction (SCLC). Band structure of $MoS_2$ calculated via density functional theory (DFT) (blue line) on a highly symmetric path along the 2D Brillouin zone for two different surface-adsorbed Ag concentrations: c) $4x10^{13}cm^{-2}$ and d) $25x10^{13}cm^{-2}$, respectively. The band structure of pristine $MoS_2$ is also included in red for comparison. Note that the band structure is shifted to coincide with the conduction band of pristine $MoS_2$.

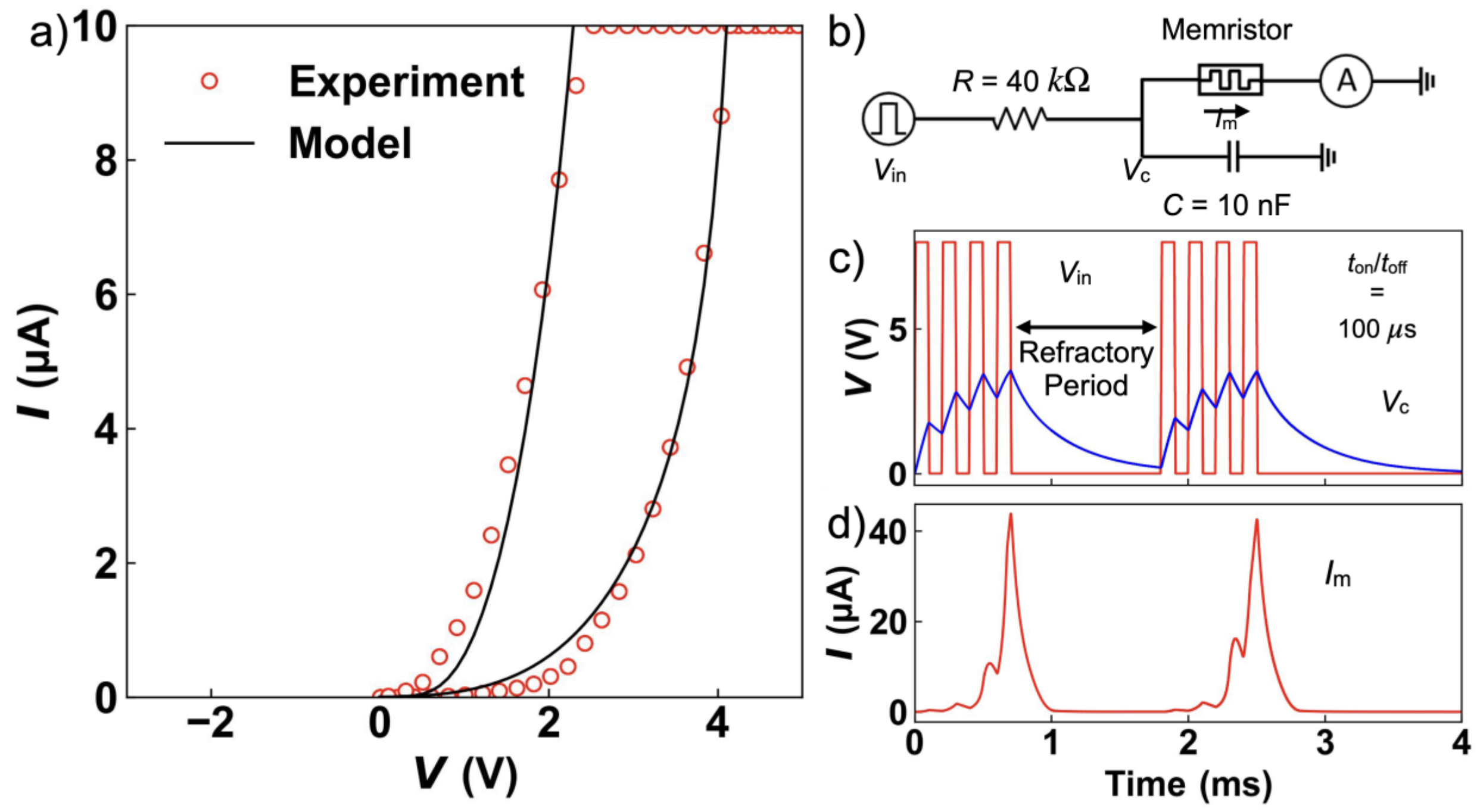

**Figure 5.** a) Experimental *I*–*V* characteristics of the experimental memristor (symbols) matched with the developed model (solid lines), showing good agreement. b) Circuit implementation of the leaky integrate-and-fire model of the neuron using the memristor. SPICE simulation results showing c) Input pulse train ($V_{in}$) with $t_{on}$ and $t_{off}$ = 100 µs and a refractory period of 1000 µs along with the voltage across the capacitor ($V_c$) and d) current through the memristor ($I_m$) showing high current spikes at the end of the integration period, emulating the firing of the neuron.

# Supporting Information

# Volatile $MoS_2$ Memristors with Lateral Silver Ion Migration for Artificial Neuron Applications

Sofia Cruces[1], Mohit D. Ganeriwala[2], Jimin Lee[1], Lukas Völkel[1], Dennis Braun[1], Annika Grundmann[3],

Ke Ran[4,5,8], Enrique G. Marín[2], Holger Kalisch[3], Michael Heuken[3,6], Andrei Vescan[3], Joachim Mayer[4,5], Andrés Godoy[2], Alwin Daus[1,7,*] and Max C. Lemme[1,8*].

[1] Chair of Electronic Devices, RWTH Aachen University, Otto-Blumenthal-Str. 25, 52074 Aachen, Germany

[2] Department of Electronics and Computer Science, Universidad de Granada, Avenida de la Fuente Nueva S/N 18071, Granada, Spain

[3] Compound Semiconductor Technology, RWTH Aachen University, Sommerfeldstr. 18, 52074 Aachen, Germany

[4] Central Facility for Electron Microscopy, RWTH Aachen University, Ahornstr. 55, 52074, Aachen, Germany

[5] Ernst Ruska-Centre for Microscopy and Spectroscopy with Electrons (ER-C 2), Forschungszentrum Jülich GmbH, Wilhelm-Johnen-Str., 52425 Jülich, Germany

[6] AIXTRON SE, Dornkaulstr. 2, 52134 Herzogenrath, Germany

[7] Sensors Laboratory, Department of Microsystems Engineering, Georges-Köhler-Allee 103, 79110 Freiburg, Germany

[8] AMO GmbH, Advanced Microelectronic Center Aachen, Otto-Blumenthal-Str. 25, 52074 Aachen, Germany

**Section S1: Proof of silver (Ag) tarnishing**

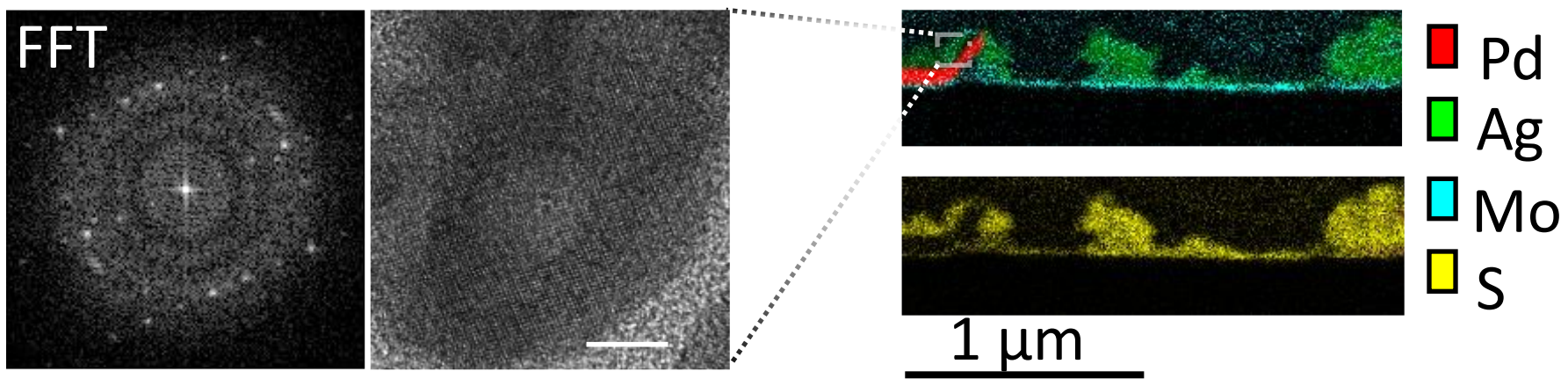

Figure S1: Proof of Ag tarnishing on a device after multiple electrical measurements. Based on the energy-dispersive x-ray spectroscopy (EDX) elemental mapping, both Ag and sulfur (S) are detected simultaneously between the palladium (Pd) and Ag metal contacts. Close to the Pd electrode region, high resolution transmission electron microscopy (HRTEM) image was recorded, and the corresponding fast Fourier transform (FFT) suggest $Ag_2S$ along [210] direction.

**Section S2: Raman mapping of the peak intensities after fabrication**

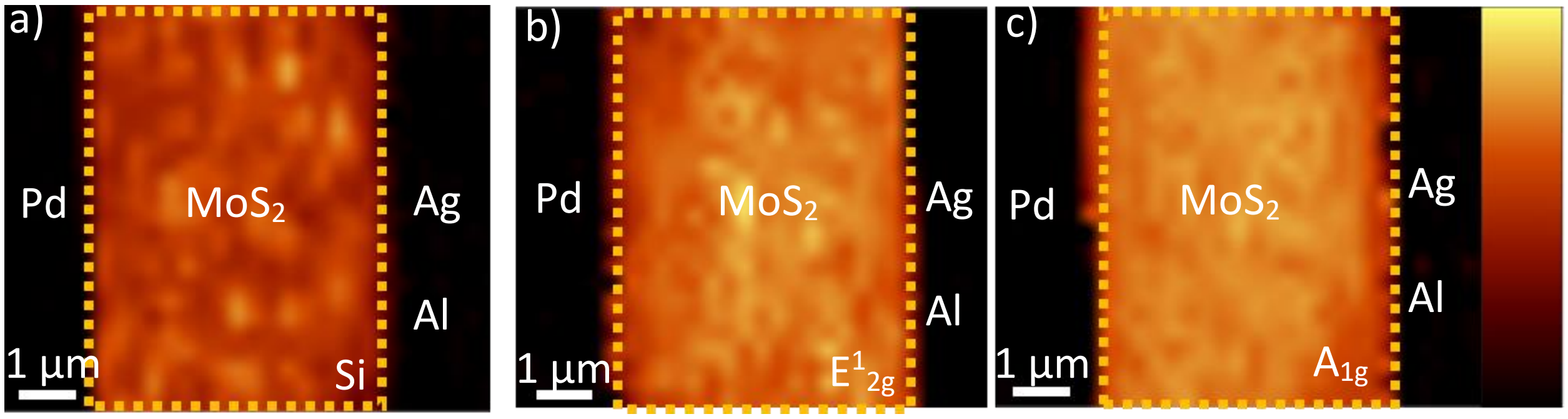

Figure S2: Raman map of the peak intensities after fabrication of the a) Si, b) $E^1_{2g}$, and c) $A_{1g}$ peaks. The integrity of the molybdenum disulfide ($MoS_2$) channels after patterning by $CF_4/O_2$ reactive ion etching was verified via Raman mapping. The color scale increases to 30 CCD counts (higher – brighter).

**Section S3: TEM characterization of $MoS_2$ after fabrication and electrical measurements**

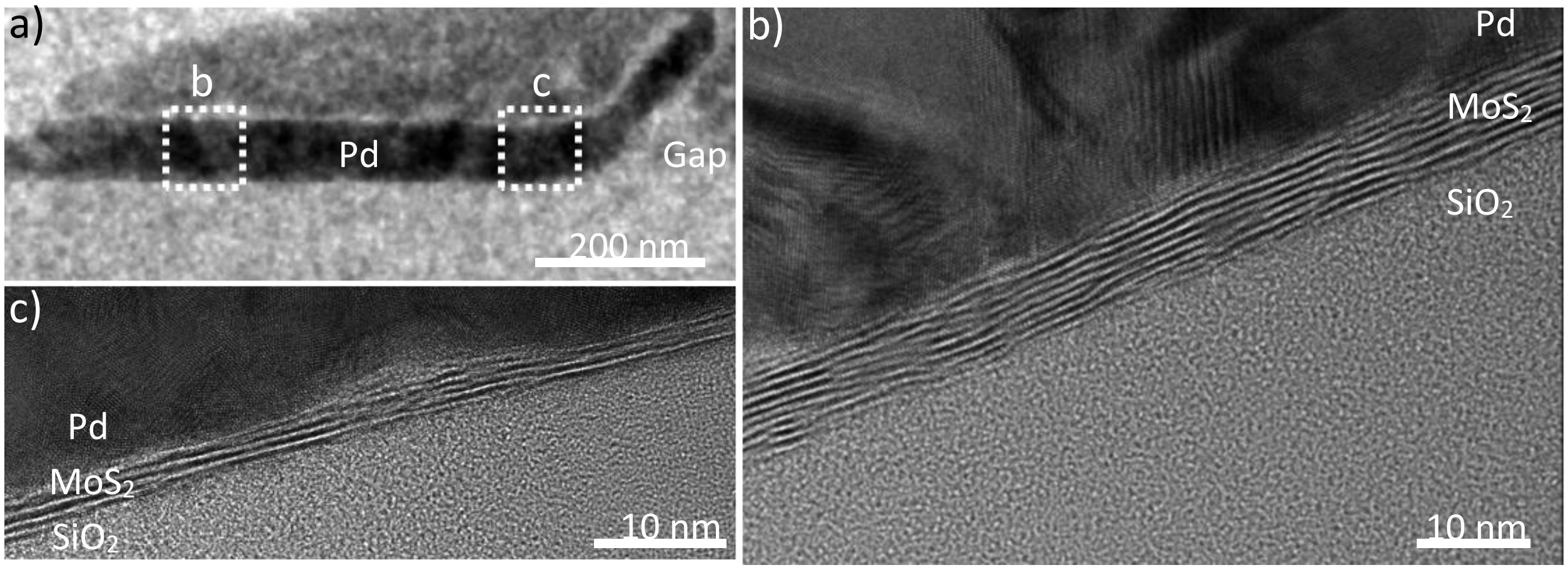

Figure S3: TEM and HRTEM cross-sectional images of $MoS_2$ after fabrication and electrical measurements. a) TEM cross-sectional image taken around the Pd electrode. Two different positions (b and c) along the Pd electrode are marked in white. b) HRTEM image of position b, which is located away from the gap between the metal electrodes. c) HRTEM image at position c, which is close to the gap between the metal electrodes. TEM imaging allows us to confirm the thickness variation of $MoS_2$ and the small difference in topography. Approaching the gap between the metal electrodes (position c), the $MoS_2$ layers were thinner, and more defects were observed. Close to the gap between the electrodes (position c), the number of layers measured was between 4 and 7, but 7 to 12 layers could be distinguished from it (position b). Some of this difference could be due to the wet transfer process, in which material can be left on the growth substrate.

**Section S4: Raman spectra of as-grown $MoS_2$ on sapphire and after transfer**

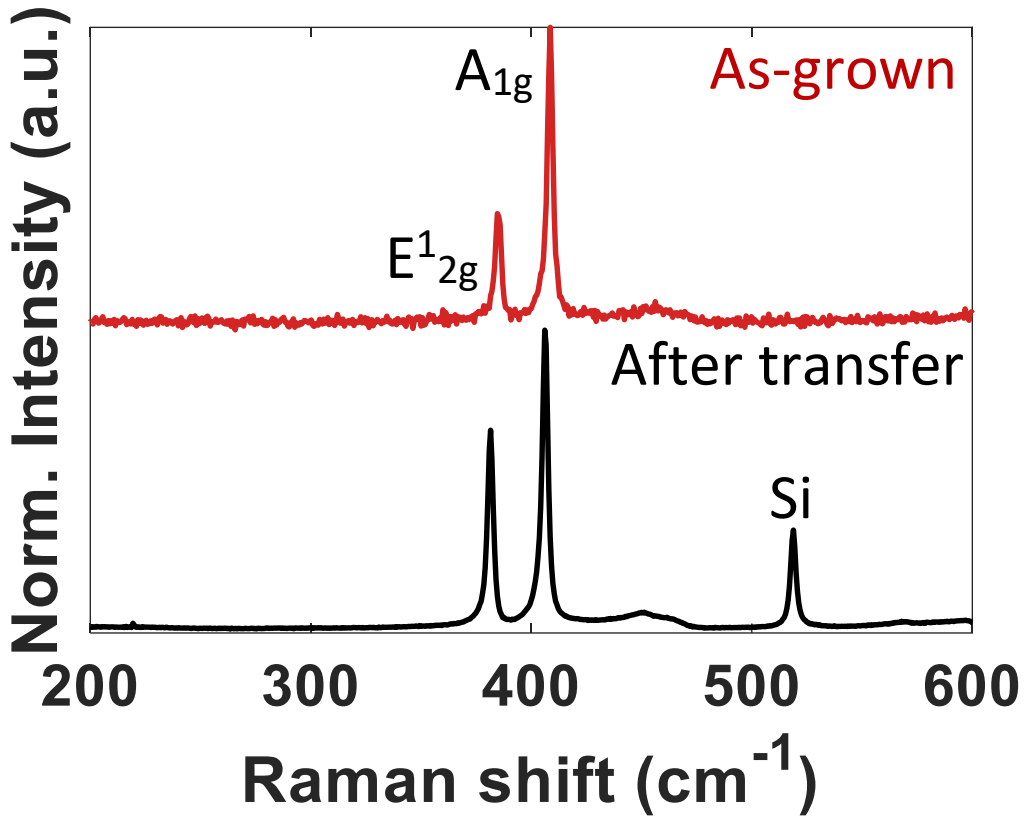

Figure S4: Raman spectra of as-grown $MoS_2$ on sapphire and after transfer on an $SiO_2$/Si substrate. Notably, the measurements were not taken at the exact same location, which could lead to a slight frequency difference due to thickness variation in the sample [1]. The extracted peaks for the as-grown $MoS_2$ on sapphire were 384.6 $cm^{-1}$ and 408.5 $cm^{-1}$ for the $E^1_{2g}$ and $A_{1g}$ peaks, respectively. These peaks coincide with those taken for more than four layers or bulk material. In the case of $MoS_2$ after transfer onto an $SiO_2$/Si substrate, the obtained values were 381.5 $cm^{-1}$ and 406.1 $cm^{-1}$ for the $E^1_{2g}$ and $A_{1g}$ peaks, respectively. These values match those previously reported for four layers or bulk [1,2].

## Section S5: Higher current compliance for volatile RS

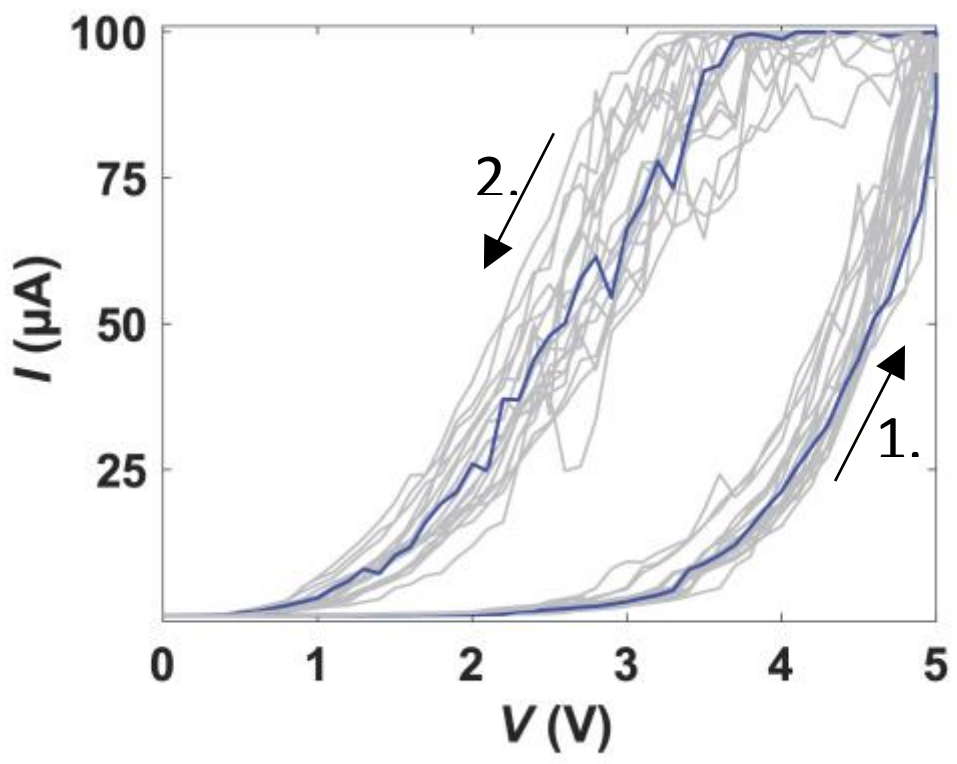

Figure S5: Twenty consecutive volatile RS cycles on a 1.2 μm device with 100 μA as the current compliance (CC). Arrows 1 and 2 show the voltage sweep direction.

## Section S6: Proof of switching for different metal combinations with and without $MoS_2$

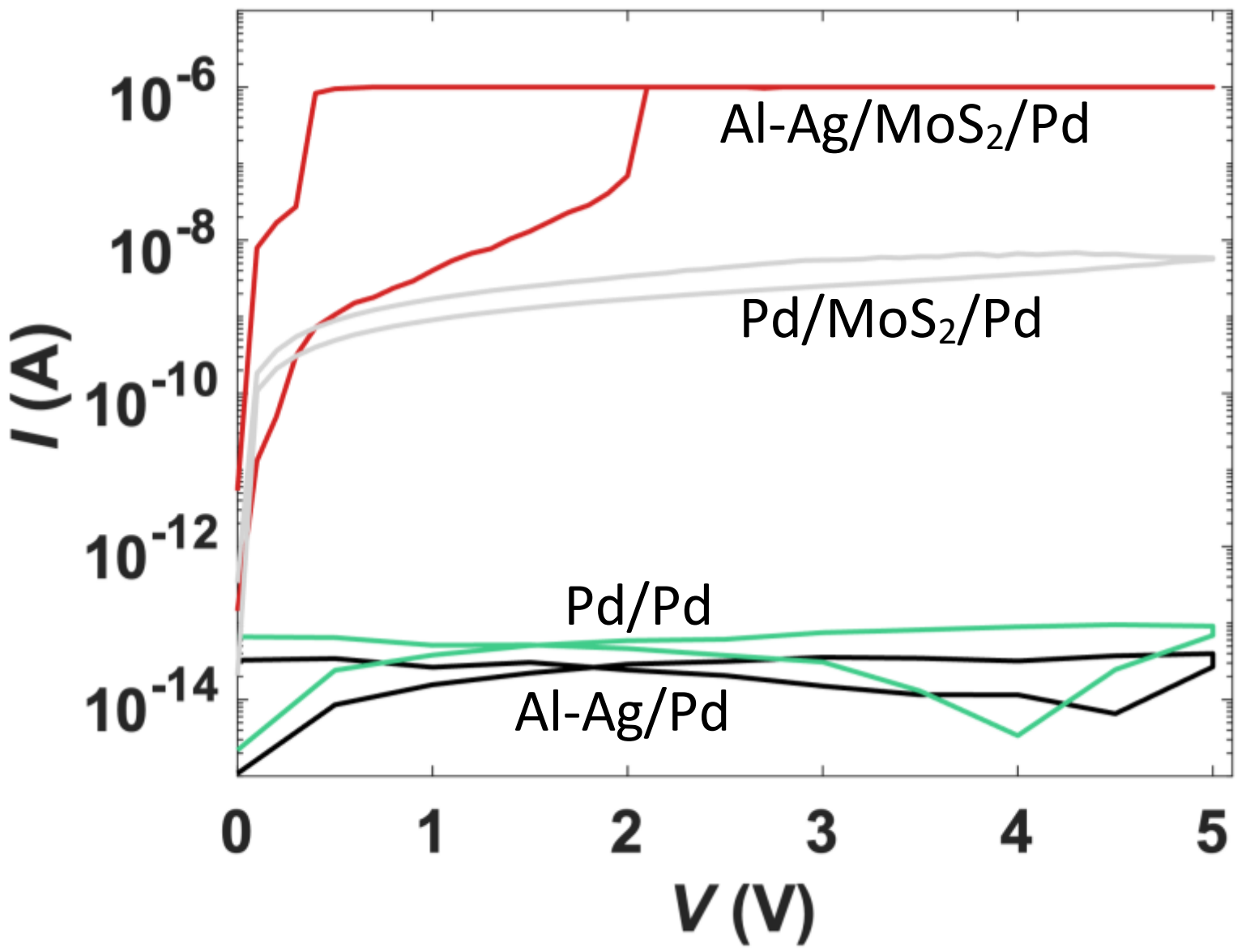

Figure S6: Current–voltage (I–V) sweeps of four devices with different metal combinations with and without $MoS_2$. These results also support that resistive switching (RS) originates from Ag ion migration on $MoS_2$. Pd/Pd- and Al-Ag/Pd-labeled measurements are electrodes where there is no $MoS_2$ bridging the gap.

**Section S7: Histogram and Gaussian fit of the device data from Fig. 2b**

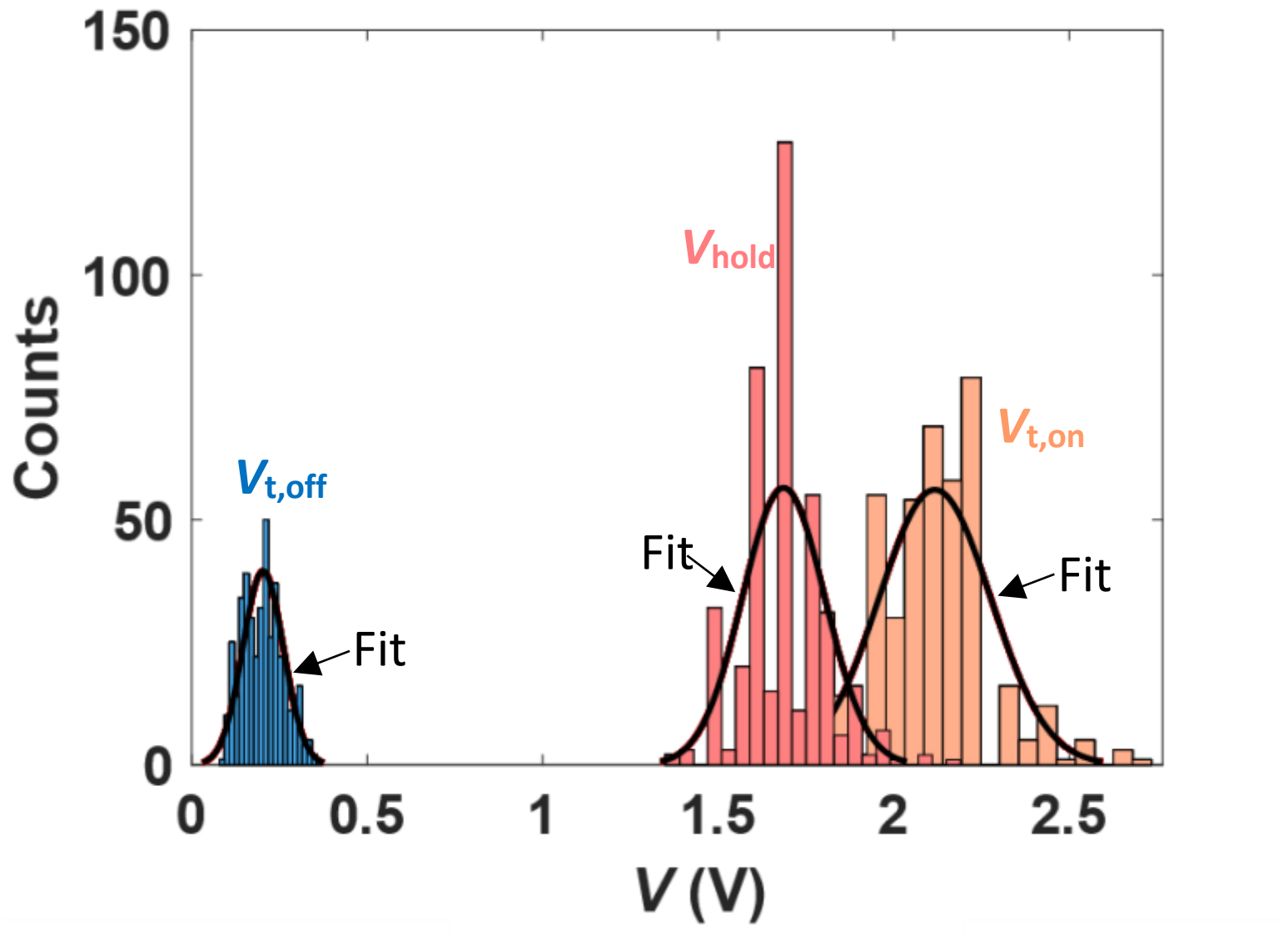

Figure S7: Histogram plot and Gaussian fit of $V_{t,on}$, $V_{hold}$, and $V_{t,off}$ extracted from endurance measurements with 416 subsequent *I*–*V* sweeps. The calculated means $V_{t,on}$, $V_{hold}$, and $V_{t,off}$ of 2.1 ± 0.1 V, 1.7 ± 0.1 V, and 0.2 ± 0.05 V, respectively, were extracted by fitting Gaussian distributions to the histogram data. A standard deviation of 0.1 V or lower indicates low cycle-to-cycle variability.

**Section S8: Subsequent *I–V* sweeps for different gap sizes**

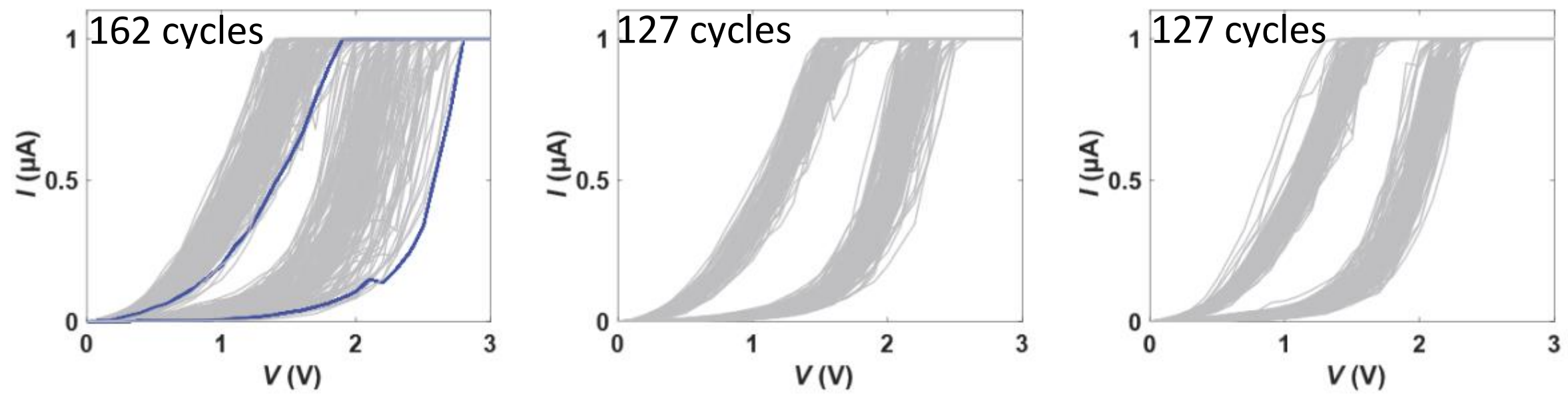

Figure S8-1: Raw data from 416 subsequent *I–V* sweeps from the device with a ~1.2 µm gap size. The first sweep is marked in blue.

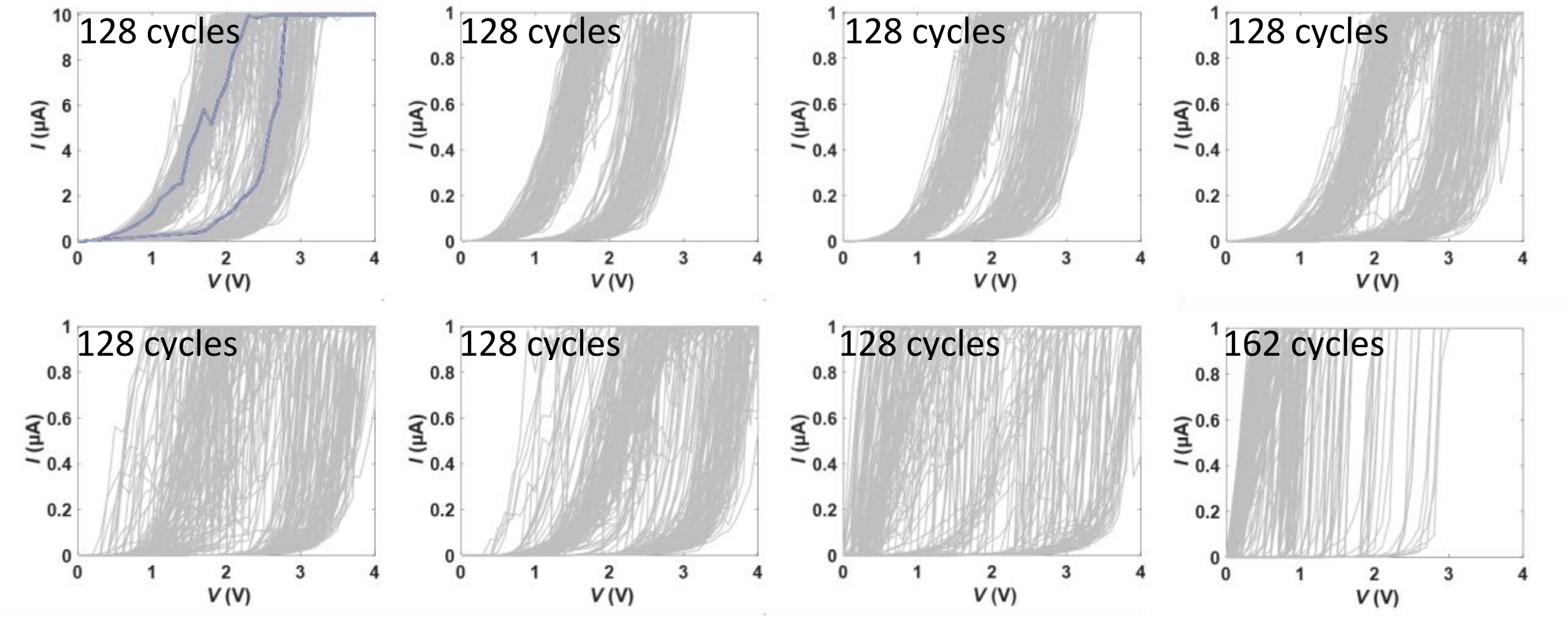

Figure S8-2: Raw data from 1058 subsequent *I-V* sweeps from the device with a ~2.1 µm gap size. The first sweep is marked in blue.

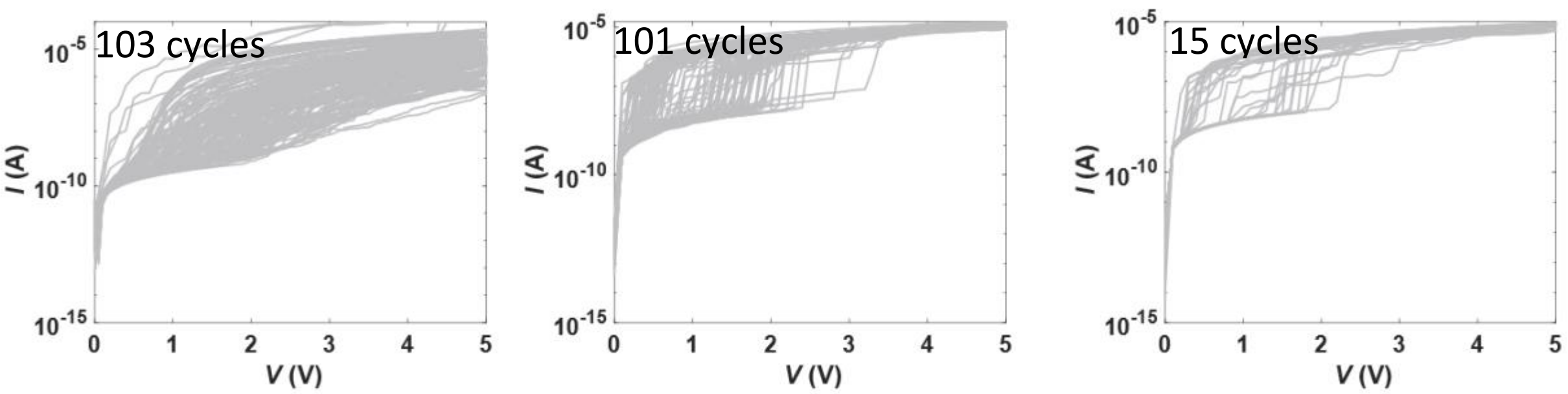

Figure S8-3: Raw data from 219 subsequent *I–V* sweeps from the device with a ~5.8 µm gap size.

The forming step was not included in the plot.

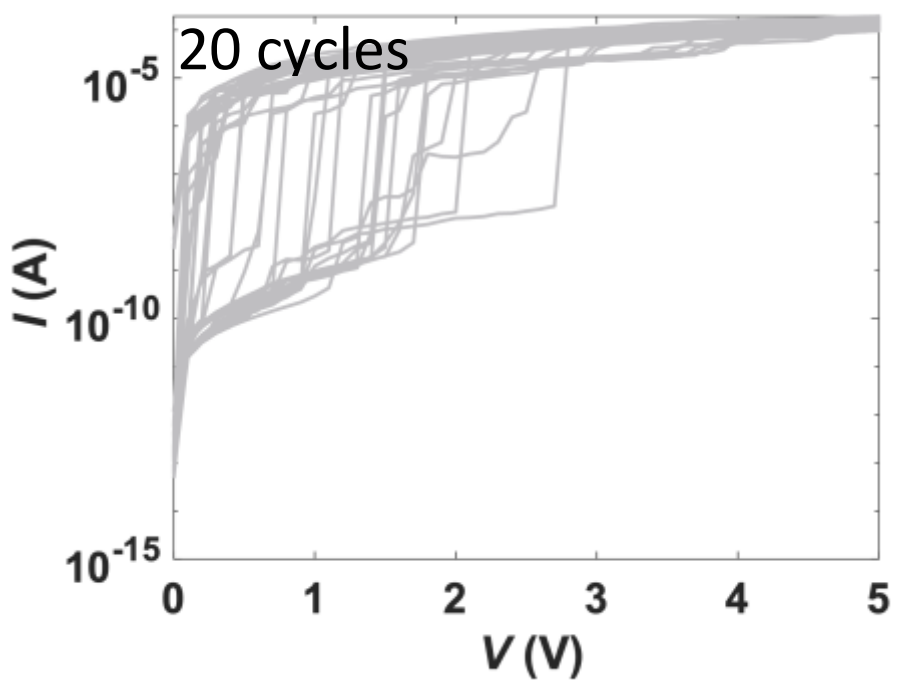

Figure S8-4: Raw data from 20 subsequent *I–V* sweeps from the device with a ~4.7 µm gap size.

The forming step was not included in the plot.

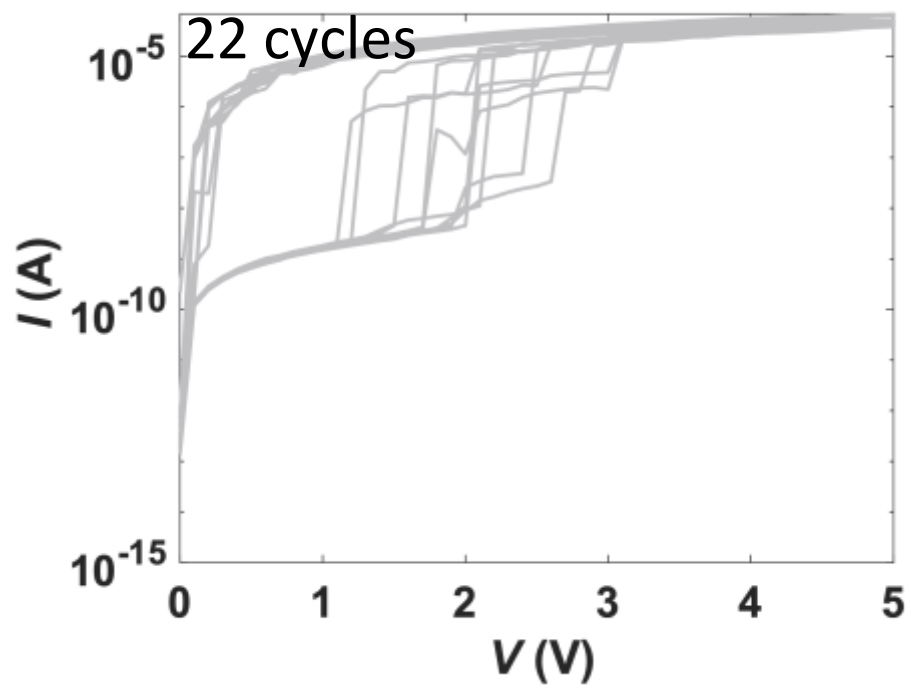

Figure S8-5: Raw data from 22 subsequent *I–V* sweeps from the device with a ~5.1 µm gap size.

The forming step was not included in the plot.

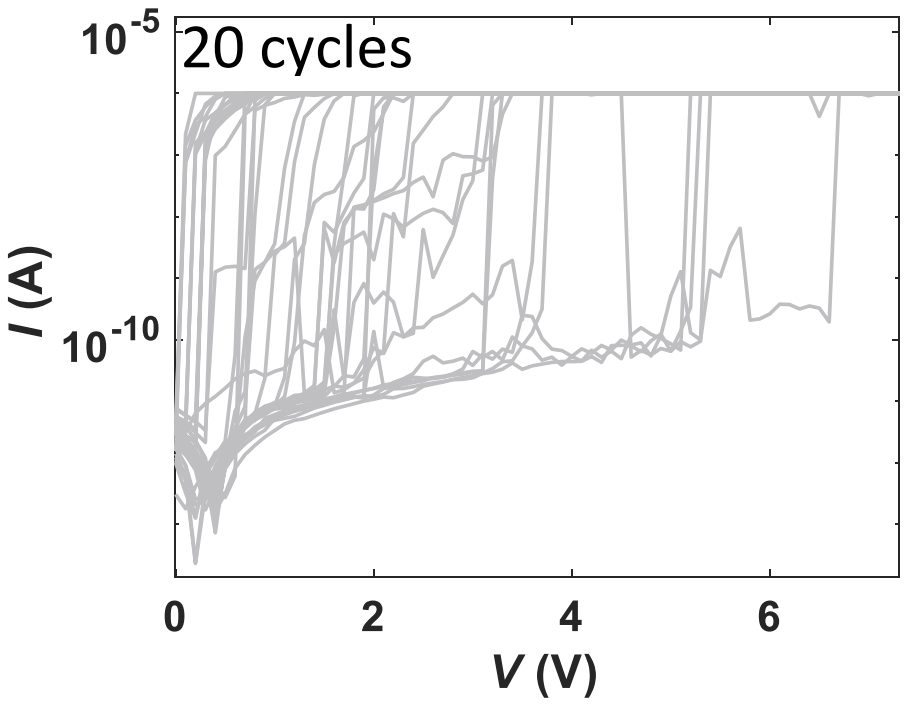

Figure S8-5: Raw data from 20 subsequent *I*–*V* sweeps from the device with a ~3.9 µm gap size.

The forming step was not included in the plot.

**Section S9: Parameter extraction for pulse measurements**

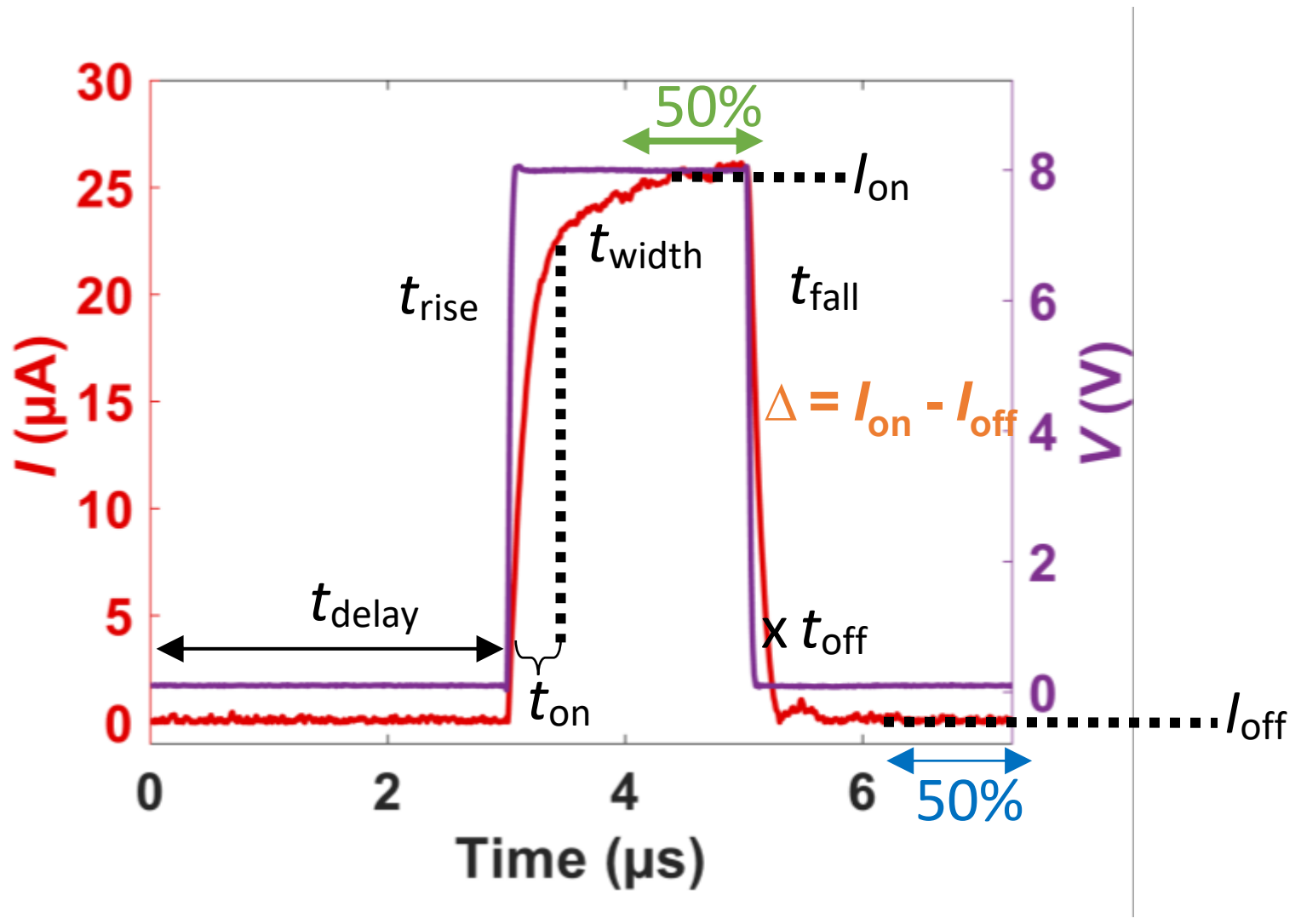

Figure S9: Pulsed waveforms with the following parameters were used: 3 µs delay time ($t_{delay}$), 50 ns rise and fall time ($t_{rise}$ and $t_{fall}$), and 2 µs width ($t_{width}$).

Extraction and calculation of the on-switching characteristics:

1. Extrapolation of data, saving the new variables.
2. Consider the second half of the switching pulse width (marked in green).
3. Calculate the average of the current values in the green region to obtain $I_{on}$.
4. Calculate the 90% of the $I_{on}$ and save this variable as $I_{on,90\%}$.
5. Find the corresponding index in x-axis and get the time for that current value.
6. Subtract $t_{delay}$ from the obtained time in step 5 and save the new variable as $t_{on}$.
7. Calculate the average of the voltage values in the green region to obtain $V_{on}$.
8. The ON-state resistance was calculated with this formula: $R_{on} = V_{on} / I_{on}$.

Extraction and calculation of the off-switching characteristics:

1. Extrapolation of data, saving the new variables.

2. Consider the second half of the base pulse width (marked in blue).
3. Calculate the average of the current values in the blue region to obtain $I_{off}$.
4. Calculate the difference (delta, Δ) between the $I_{on}$ and $I_{off}$.
5. Add the 10% of the delta to the $I_{off}$: $I_{off,delta} = I_{off} + 10\%\Delta$.
6. Find the corresponding index in x-axis and get the time for that current value.
7. Subtract ($t_{delay}$+ $t_{rise}$+ $t_{width}$) from the obtained time in step 6 and save the new variable as $t_{off}$ ($t_{relax}$).
8. Calculate the average of the voltage values in the blue region to obtain $V_{off}$.
9. The OFF-state resistance was calculated with this formula: $R_{off} = V_{off} / I_{off}$.

**Section S10: Comparison of device characteristics with those of other similar 2D-based memristive devices reported in the literature**

| Ref. | Structure | Growth | No. Layers | Gap length | Forming | Volatile | Cycles | $V_{t,on}$ (V) |
|---|---|---|---|---|---|---|---|---|
| **This work** | **Planar** | **MOCVD** | **7-12** | **1.2 µm** | **No** | **Yes** | **1058** | **~2** |
| [3] | Planar | CVD | 1 | 2-7.5 µm | Yes | No | 12 | 3.5-8.3 |
| [4] | Planar | CVD | 1 | 5–15 µm | Yes | No | 475 | ~20 |
| [5] | Planar | Exfoliated | 5 | 10-40 nm | Yes | Yes | 300 | ~0.4 |
| [6] | Planar | Exfoliated | 4 | 250 nm | Yes | No | 1 | ~2 |
| [7] | Planar | Exfoliated | ~13 | ~4 µm | N/A | No | 100 | ~10 |

Table S1: Benchmarking table of publications showing different characteristics of the devices.

**Section S11: Pulse programming conditions for $R_{on}$, $t_{on}$, and $t_{off}$ extraction**

| Voltage (V) | Rise | Width | Fall | Base Voltage (V) |
|---|---|---|---|---|
| 2 | 1 µs | 20 ms | 1 µs | 0.1 |
| 3 | 1 µs | 3 ms | 1 µs | 0.1 |
| 4 | 50 ns | 20 µs | 50 ns | 0.1 |
| 5 | 50 ns | 5 µs | 50 ns | 0.1 |

Table S2: Pulse programming conditions for $R_{on}$, $t_{on}$, and $t_{off}$ extraction.

**Section S12: First-principles calculations of Ag adsorption on the $MoS_2$ surface**

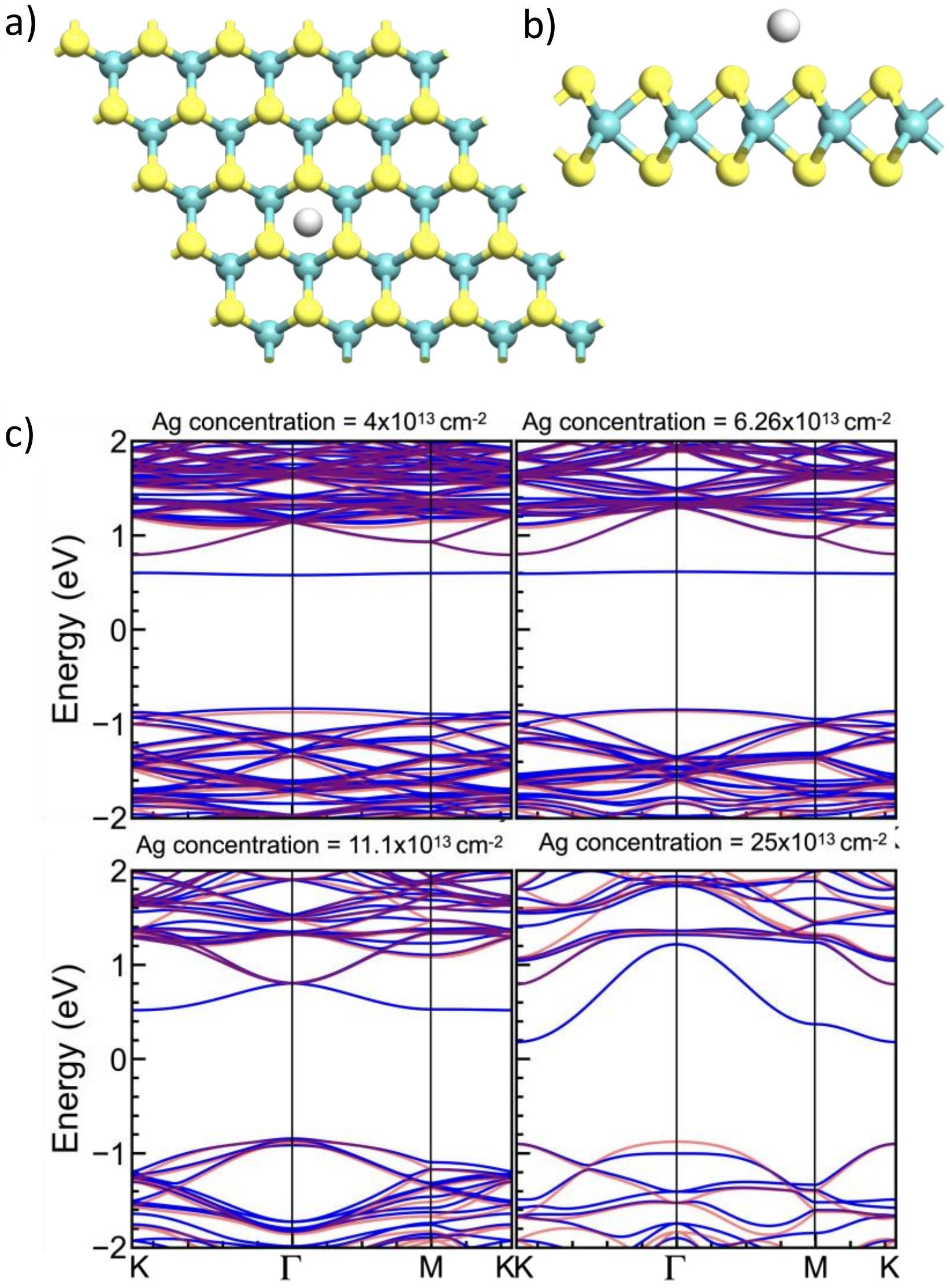

Figure S10: a) Top view and b) side view of monolayer $MoS_2$ supercell with surface-adsorbed Ag and c) band structure calculated via density functional theory (DFT) (solid blue) for surface-adsorbed Ag with gradually increasing concentration, showing the transition of the localized to delocalized gap state. The band structure of pristine $MoS_2$ is also shown in semitransparent red for comparison.

**Section S13: Physical modeling of the volatile resistive switching (RS) memristor**

According to the *ab initio* results, the presence of Ag atoms along the $MoS_2$ channel and their concentration determine the electronic transport mechanism in the memristive device. Figure S10 shows a schematic illustration of the analyzed structure, where *L* is the distance between the Pd and Ag contacts and *d* represents the average distance among Ag ions.

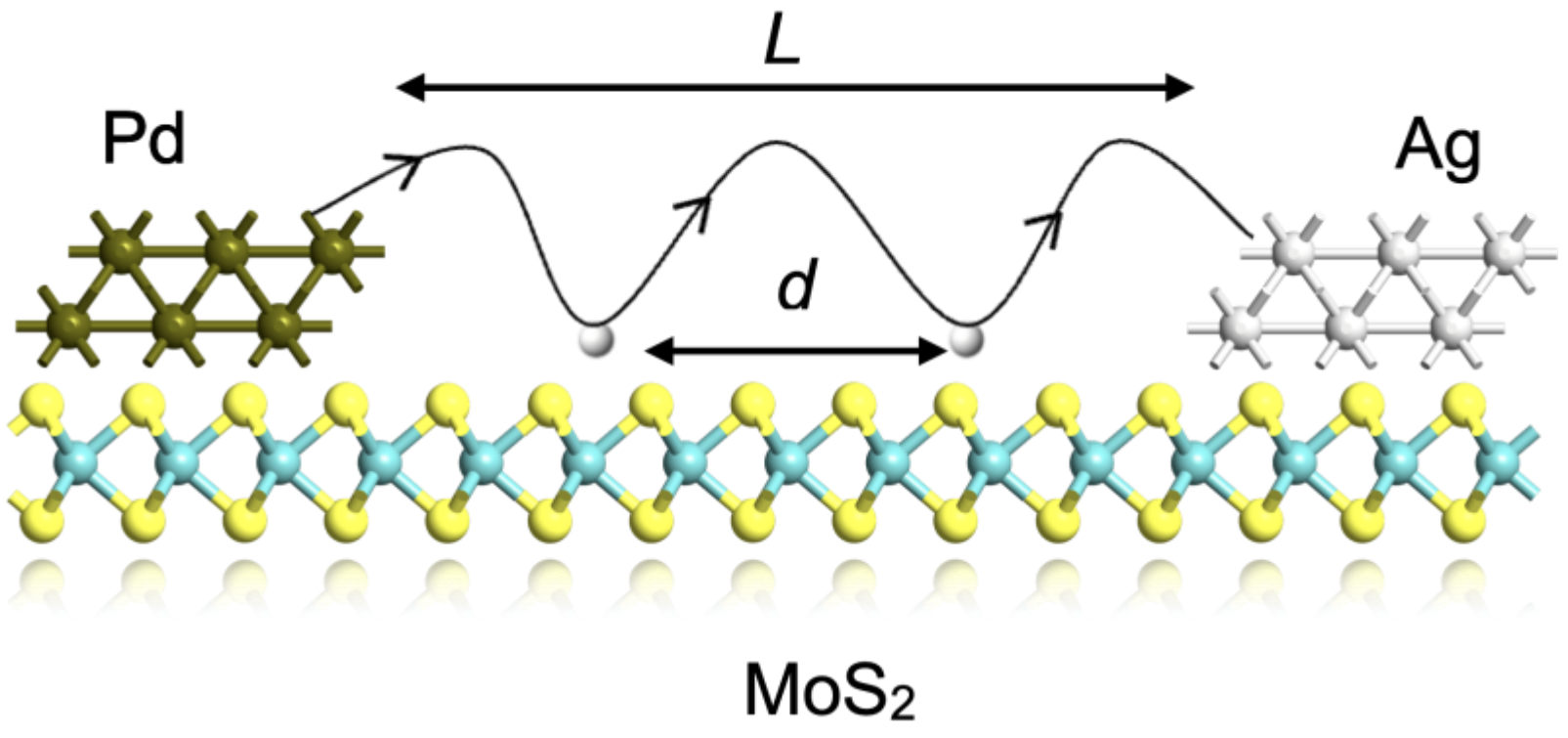

Figure S11: Schematic showing the monolayer $MoS_2$ with surface-adsorbed Ag, where *L* is the gap length and d is the average distance between adjacent Ag atoms.

As discussed in the main text, the I–V characteristics of the HRS and LRS can be attributed to hopping and space-charge-limited conduction (SCLC) transport, respectively.

The complete memristor model is developed in two stages: i) first, the movement of the Ag ions within the device is considered, and ii) this Ag dynamics is connected to the current conduction mechanism via electron transport.

To this end, the movement of the Ag ions is modeled on the basis of the probability of the ions overcoming the migration barrier according to the Arrhenius law. The average distance between adjacent Ag ions (*d*) is considered the state variable, defining the dynamics of the resistance state. Following, ref.[9] The time derivative of *d* can be written as

$$\frac{d}{dt}(d) = -V_0 exp\left(-\frac{E_a}{kT}\right) sinh\left(\gamma \frac{qA_0}{kT} \frac{V - SHIFT}{L}\right)$$

$$\gamma = \gamma_0 - \beta d^3$$

where $E_a$ is the migration barrier for Ag ions; *k* is the Boltzmann constant; *T* is the absolute temperature; *q* is the elementary charge; *L* is the gap length (distance between two contacts); and $V_0$, $A_0$, SHIFT, $\gamma_0$ and $\beta$ are fitting parameters. Here, the introduction of the parameter SHIFT ensures that the memristor resets when the voltage value crosses the off voltage ($V_{t,off}$), capturing the volatile nature of the memristor.

The value of *d* determines the conduction state of the memristor, as it depends on the history of the applied voltage. It is bounded in the model implementation to avoid the inclusion of unphysical values. The current conduction due to electron transport is then modeled as follows:

From Ohms' law, the electron current in the memristor can be written as

$$I = nq\mu A \frac{V}{L}$$

where *n* is the electron density, *A* is the cross-sectional area and $\mu$ is the mobility. In the case of disordered systems, the mobility due to hopping is given by [10]:

$$\mu = \mu_0 \exp\left(-\frac{E_{act}}{kT} + \frac{G_0}{kT}\sqrt{\frac{V}{L}}\right)$$

where $\mu_0$ is the field-independent mobility, $E_{act}$ is the activation energy for electron hopping and $G_0$ is the field enhancement factor, which is used here as a fitting parameter.

With this definition of $\mu$, the current can be written as

$$I = I_0 \cdot n \cdot exp\left(\frac{G_0}{kT}\sqrt{\frac{V}{L}}\right) \cdot \frac{V}{L}$$

$$I_0 = q\mu_0 A exp\left(-\frac{E_{act}}{kT}\right)$$

In addition to the carrier hopping assisted by the applied field captured through the $\mu$, the distance between the Ag ions is also dynamically modified with the applied bias. This modification affects the band structure of $MoS_2$, as confirmed via DFT calculations (Fig. S10). This change in the band structure can be captured through the electron density (*n*), which can be modeled on the basis of *d* as follows:

**Case 1:** When d → 0, a delocalized band appears in the bandgap, giving rise to band transport, and the electron density in such a case ($n_0$) can be written via the Gauss law as

$$q n_0 L = \epsilon_s \frac{V}{L}$$

$$\therefore n_0 = \frac{\epsilon_s}{q}\frac{V}{L^2}$$

where $\epsilon_s$ is the permittivity of $MoS_2$.

**Case 2:** When *d* → *L*, the electron density is equal to the number of electrons at the localized trap state ($n_t$)

Therefore, at any intermediate *d*, the value of *n* will be the addition of the trap carrier $n_t$ and the fraction of $n_0$ that can jump overcoming the barrier, which is proportional to the jump rate $exp(-\alpha d)$ [11]. The total *n*, therefore, is modeled as

$$n = \left(\frac{n_0 exp(-\alpha d) + n_t}{1 + exp(-\alpha d)}\right)$$

Using the derived value of *n*, the total electron current through the memristor can be written as

$$I = I_0 \left[\frac{\frac{\epsilon_s V}{qL^2} exp(-\alpha d) + n_t}{1 + exp(-\alpha d)}\right] exp\left(\frac{G_0}{kT}\sqrt{\frac{V}{L}}\right)\frac{V}{L}$$

The expression for the current can be simplified in the following scenarios:

1: when *d* is large,

$$I = I_0 n_t exp\left(\frac{G_0}{kT}\sqrt{\frac{V}{L}}\right)\frac{V}{L}$$

which describes hopping transport with field-enhanced mobility [10] or Poole-Frenkel (PF)-type transport that follows the experimentally observed behavior in the HRS, i.e., $\frac{I}{V} \propto exp(\sqrt{V})$.

2: when *d* is small,

$$I = I_0 \frac{\epsilon_s}{q} \exp\left(\frac{G_0}{kT}\sqrt{\frac{V}{L}}\right)\frac{V^2}{L^3}$$

which describes the SCLC with PF-enhanced mobility [12] and follows the experimentally observed behavior in the LRS.

Therefore, the model proposed here efficiently captures the switching mechanism as the carrier transport changes from hopping to SCLC via a single continuous equation.